\documentstyle[PASJadd,psfig]{PASJ95}
\draft
\begin{document}

\title{Chemical Evolution in the Large Magellanic Cloud}

\author{Shigehiro {\sc Nagataki} and Gentaro {\sc Watanabe}
\\[12pt]
{\it Department of Physics, School of Science, the University
of Tokyo, 7-3-1 Hongo, Bunkyoku, Tokyo 113 }\\
{\it E-mail(TY): Nagataki@utaphp5.phys.s.u-tokyo.ac.jp}
}

\abst{
We present a new input parameter set of the Pagel model (Pagel \&
Tautvai$\rm \breve{s}$ien$\rm \dot{e}$ 1998) for the Large
Magellanic Cloud (LMC) in order to reproduce the
observations, including the star formation rate (SFR) history.
It is concluded that the
probability for (3-8)$M_{\odot}$ stars to explode as SNe Ia
has to be quite high ($\sim  0.17$) in the LMC.
As a result, a steep initial mass function (IMF) slope and existence of 
the outflow are not needed in order to attain the low [O/Fe] ratio in the LMC.
As for the current supernova ratio, a high ratio ($\sim 1.3$) is
concluded by the new parameter set, which is consistent
with the recent X-ray observations.
}

\kword{galaxies: chemical evolution ---  Magellanic Clouds --- 
Large Magellanic Cloud --- stars: abundances --- supernovae: general}

\maketitle
\thispagestyle{headings}

\section
{Introduction}

Chemical evolution model for a galaxy is a tool from which we can infer the
history of the galaxy. A chemical evolution model will be believed to be
right if it can reproduce the observations about the galaxy, such as
the age-metallicity relation, the present Fe distribution of stars,
and the [O/Fe] versus [Fe/H] diagram. As a result, we can infer some
important informations about the events in the galaxy, such as the
inflow rate of the material and the
chemical compositions of Type Ia/II supernovae (SNe Ia/SNe II)
using such a 'right'
chemical evolution model. These informations give some constraints on 
other theories and numerical calculations.
For example, people who are devoted to the formation of a galaxy or the
nucleosynthesis in a supernova should take these constraints into
consideration.
If they can meet them, it means that their calculations are supported by the
chemical evolution model. This is the reason why we try to construct a 
good chemical evolution model.

There are a lot of parameters in a chemical evolution model.
We can not know $a$ $priori$ whether the number of the parameters is too 
much or too little. One concerned with a chemical evolution model
should try to reduce the number of the free parameters in his model,
while he should try to reproduce the observational data as many as
possible. Unless such an effort, degenerate solutions that
can explain some 'selected' observations will be reported one after
another to the world.
Even worse, if the history of the galaxy can not be determined
uniquely using these solutions,
we can not extract important informations about the events in
the galaxy. This means that the aim of the chemical evolution model
is not accomplished at all.

In this study, we will consider the chemical evolution in the
Large Magellanic Cloud (LMC). There are some excellent works about it.
For example, Tsujimoto et al (1995) and Pagel and Tautvai$\rm
\breve{s}$ien$\rm \dot{e}$ (1998) reproduce the age-metallicity
relation, the [O/Fe]-[Fe/H] diagram, and the present mass fraction of
gas very well. However, some of the conclusions presented in Tsujimoto
et al (1995) are incompatible with those in Pagel and Tautvai$\rm
\breve{s}$ien$\rm \dot{e}$ (1998). The Tsujimoto model predicts (i) a
steeper initial mass function (IMF) slope ($\sim$ 1.55 -- 1.72)
compared to that in our galaxy ($\sim$ 1.35; Salpeter 1955) and (ii) a
higher relative frequency of Type Ia to Type II supernovae ($N_{\rm
Ia}/N_{\rm II} \sim$ 0.2 -- 0.25) compared to that in our galaxy
(0.15; Tsujimoto et al. 1995). On the other hand, the Pagel model
does not require (i) a steeper IMF slope and does not predict (ii) a
higher $N_{\rm Ia}/N_{\rm II}$ compared to those in our galaxy.
The concluded histories of the star formation and the inflow rate of
the material are also different between their models. This means that their
solutions are the degenerate ones mentioned above.
This situation is contrast to that for our galaxy. Their solutions for 
our galaxy resemble each other (Tsujimoto et al. 1995; Pagel \&
Tautvai$\rm \breve{s}$ien$\rm \dot{e}$ 1995).

It is true that we can not always obtain
an unique solution even if we try to. Various solutions will be allowed 
when no strict constraint is derived from the observations. In fact,
the number of the observations for the LMC is fewer than that
for our galaxy. However, we think they do not make full use of the
observational data for the LMC. For example, observations of
the present Fe distribution of stars which will reflect the star
formation rate (SFR) history are used to determine the parameters for
the model of our galaxy (Tsujimoto et al. 1995; Pagel \& Tautvai$\rm
\breve{s}$ien$\rm \dot{e}$ 1995), while
such observations are not used for the model of the LMC
(Tsujimoto et al. 1995; Pagel \& Tautvai$\rm \breve{s}$ien$\rm \dot{e}$
1998). Even worse, their works contain
trivial inconsistency in their analysis (see subsection 2.2 for details).
That is why we should obtain
a solution for the LMC using the observations of the SFR 
and a consistent analysis before we conclude that the history of the
LMC can not be determined uniquely by a chemical evolution model.

In this paper, we use the Pagel model and determine the values of 
the free parameters in his model by the observations of the LMC,
including the observation of the SFR history (Olszewski 1993).
We will discuss the meaning of the predictions derived from the model
doing a consistent analysis.
In section 2, formulation of the Pagel model is explained.
Results are shown in section 3. Summary and discussion
are presented in section 4.

\section{Formulations} \label{model}
\subsection{The Pagel Model}\label{pagel}
\indent

In this subsection, we present the formulation of the Pagel model.
We consider one zone and assume that the gas is distributed uniformly
and the heavy elements are well-mixed within the zone. Once we make
such an assumption, any physical quantum is determined uniquely as a
function of time in the model. This means that we do not try to
reproduce the dispersions of the observations from the beginning. The
aim of such a model is to reproduce the averaged value of the observations.
If it can not, it will lose its justification for existence.

They introduced a dimensionless time-like variable $u$, defined by
\begin{eqnarray}
u \equiv \int^{t}_{0} \omega(t')dt'
\label{eqn1}
\end{eqnarray}
where $\omega(t)$ is the transition probability for diffuse material
('gas') to change into stars in unit time at time $t$. This means
the simple linear star formation law
\begin{eqnarray}
\frac{\rm d \it s}{\rm d \it t} = \omega g \;\;\;\ \rm or \;\;\; \it
\frac{\rm d \it s}{\rm d \it u} = g 
\label{eqn2}
\end{eqnarray}
where $s(u) = \alpha S(u)$ is the mass in the form of long-lived stars
(including compact remnants), and $S$ is the mass of all stars that
were ever born up to time $t$. $\alpha$ is the lock-up fraction
(assumed constant). The fate of a star is divided into three classes;
a long-lived star, Type Ia supernova, and Type II supernova. The ratio 
of their occurrence possibility is assumed to be
$\alpha$:$\alpha_1$:1-$\alpha$-$\alpha_1$. $g(u)$ is the mass of gas
in the system under consideration.

The evolution of the total baryon mass of the system is given by
\begin{eqnarray}
\frac{\rm d}{\rm d \it t}(s + g) = f(t) + e(t)
\label{eqn22}
\end{eqnarray}
where $f(t)$ and $e(t)$ are the inflow and outflow rates of the
material, respectively. Inflow is assumed to occur at a rate
\begin{eqnarray}
f(t) = \omega (t) {\rm e}^{-u}.
\label{eqn3}
\end{eqnarray}
The inflowing material is assumed to be unprocessed.
On the other hand, outflow is assumed to occur at a rate
\begin{eqnarray}
e(t) = \eta \frac{\rm d \it s}{\rm d \it t} = \eta \omega g.
\label{eqn4}
\end{eqnarray}
$\eta$ is a free parameter.

We explain the evolution of the metallicity. The mass fraction of
$i$th element in the gas, $Z_i$, is divided into two terms. One is the 
term for the instantaneous recycling, $Z_{1,i}$, and the other is the
term for the delayed production, $Z_{2,i}$. Type II/Ia supernovae
contribute to the evolution of $Z_{1,i}$ and $Z_{2,i}$, respectively.
The life-time of Type
II supernova is much shorter ($\sim$ several Myrs) than the age
of the LMC ($\sim$ 15 Gyrs). So Type II supernovae can be treated to 
explode when their progenitors are born in a chemical evolution model.
This is the meaning of the instantaneous recycling.
On the other hand, the life-time of Type Ia supernovae (few Gyrs) is
comparable to the age of the LMC. So their life-time is taken into
consideration in a chemical evolution model.  This is the meaning of
the delayed production. The evolution of $Z_{1,i}$ obeys the following
equation:
\begin{eqnarray}
\label{eqn71}
\frac{\rm d}{\rm d \it t} (gZ_{1,i})&&= \frac{\rm d \it S}{\rm d \it t} 
(1-\alpha-\alpha_1) x_i - \frac{\rm d \it S}{\rm d \it t} \alpha
Z_{1,i} \;\;\;\;\;\;\;\;
\left(x_i=\frac{m_{i, \rm II}}{M_{\rm II}}\right) \\ \nonumber
&&= \frac{\rm d \it s}{\rm d \it t}
\frac{1-\alpha-\alpha_1}{\alpha}x_i - \frac{\rm d \it s}{\rm d \it
t}Z_{1,i} \\ \nonumber
&&= \omega g P_{1,i} - \omega g Z_{1,i} 
\end{eqnarray}
where $m_{i, \rm II}$ and $M_{\rm II}$ are averaged mass of $i$th
heavy element and total mass of Type II supernova. $m_{i, \rm II}$ is
defined as
\begin{eqnarray}
m_{i, \rm II} = \frac{\int^{m_u}_{m_l}m_{i,\rm II}(m) \phi (m)
m^{\rm -1} \rm d \it m}{\int ^{m_u}_{m_l} \phi (m) m^{\rm -1} \rm d
\it m} \;\;\;\; \rm with \;\;\; \it \phi (m) = m^{-x}.
\label{eqn72}
\end{eqnarray}
$m_{i,\rm II}(m)$ is the $i$th heavy-element mass produced in a star
of main-sequence mass $m$. $\phi (m)$ is the initial mass function
(IMF) and $x$ is the slope of the IMF.
The bounds for SNe II progenitors were taken to be $m_l
= 10 M_{\odot}$ and $m_u = 50 M_{\odot}$, respectively (Tsujimoto et
al. 1995).
$M_{\rm II}$ is defined as
\begin{eqnarray}
M_{\rm II} = \frac{\int^{m_u}_{m_l} \phi (m)
 \rm d \it m}{\int ^{m_u}_{m_l} \phi (m) m^{\rm -1} \rm d
\it m}.
\label{eqn73}
\end{eqnarray}

The evolution of $Z_{2,i}$ is given by
\begin{eqnarray}
\label{eqn74}
\frac{\rm d}{\rm d \it t} (gZ_{2,i})&&= \frac{\rm d \it S \rm ( \it t
- t_{\rm Ia} \rm )}{\rm d \it t} \alpha_1 y_i - \frac{\rm d \it S}{\rm
d \it t} \alpha Z_{2,i} \;\;\;\;\;\;\;\;
\left(y_i=\frac{m_{i, \rm Ia}}{M_{\rm Ia}}\right) \\ \nonumber
&&= \frac{\rm d \it s ( \it t - t_{\rm Ia} \rm )}{\rm d \it t}
\frac{\alpha_1}{\alpha}y_i - \frac{\rm d \it s}{\rm d \it
t}Z_{2,i} \\ \nonumber    
&&= \omega (t-t_{\rm Ia}) g(t-t_{\rm Ia}) P_{2,i} - \frac{\rm d \it
s}{\rm d \it t}Z_{2,i} 
\end{eqnarray}
where $t_{\rm Ia}$ is the averaged lifetime of SNe Ia progenitors.

In their formulation, the frequency of SNe Ia ever occurred
relative to SNe II is obtained as
\begin{eqnarray}
\frac{N_{\rm Ia}}{N_{\rm II}} = \frac{P_{2,i}}{P_{1,i}}\frac{m_{i,\rm
II}}{m_{i,\rm Ia}} \frac{s(u_{\rm now} - \omega t_{\rm Ia}
)}{s(u_{\rm now})}
\label{eqn8}
\end{eqnarray}
where $u_{\rm now}$ is the present value for $u$. 
When we ignore the lifetime of supernova progenitors, $N_{\rm
Ia}/N_{\rm II}$ is also written as
\begin{eqnarray}
\frac{N_{\rm Ia}}{N_{\rm II}} = \frac{A \int_{3M_{\odot}}^{8M_{\odot}} 
m^{-(1+x)}}{\int_{10M_{\odot}}^{50M_{\odot}}m^{-(1+x)}},
\label{eqn81}
\end{eqnarray}
where $A$ is the probability for (3-8)$M_{\odot}$ stars to explode
as SNe Ia (Tsujimoto et al. 1995). So, we can estimate the value of $A$
when $N_{\rm Ia}/N_{\rm II}$ and $x$ are determined from a chemical
evolution model.

As for the current supernova rates, we can calculate from Eq.~(\ref{eqn2})
and~(\ref{eqn8}) as
\begin{eqnarray}
\frac{\dot{N_{\rm Ia}}}{\dot{N_{\rm II}}} =
\frac{P_{2,i}}{P_{1,i}}\frac{m_{i,\rm
II}}{m_{i,\rm Ia}} \frac{\omega ( \it t - t_{\rm Ia} \rm ) 
g ( \it t - t_{\rm Ia} \rm ) }{\omega g}. 
\label{eqn92}
\end{eqnarray}

We also present here another physical quantum which is used in our analysis.
That is $r$, which is the contribution of SNe Ia to the enrichment
of heavy-elements in the gas. $r$ is defined as
\begin{eqnarray}
r = \frac{\omega_{\rm Ia} M_{\rm Ia} N_{\rm Ia}}{\omega_{\rm Ia}
M_{\rm Ia} N_{\rm Ia} + \omega_{\rm II} M_{\rm II} N_{\rm II}}
\label{eqn11}
\end{eqnarray} 
where $\omega_{\rm Ia}$ and $\omega_{\rm II}$ are mass fractions of
heavy-element ejected into the interstellar gas from SNe Ia and SNe
II, respectively. $\omega_{\rm Ia}$ and $\omega_{\rm II}$ are defined
as
\begin{eqnarray}
\omega_{\rm Ia} = \frac{ c_{\rm g} g Z_{\rm g, Fe}}{c_{\rm s} s Z_{\rm 
s, Fe}
+ c_{\rm g} g Z_{\rm g, Fe} + c_{\rm out} Z_{\rm out, Fe}} \\
\omega_{\rm II} = \frac{g Z_{\rm g, O}}{ s Z_{\rm s, O} + g Z_{\rm g, 
O} + m_{\rm out} Z_{\rm out, O}}
\label{eqn12}
\end{eqnarray}
where $Z_{\rm g}$ are the heavy-element abundance in unit mass of the
gas. $Z_{\rm s, out}$ are the heavy-element abundances averaged over the
metallicity distribution of stars and outflow, respectively. $m_{\rm out}$
is the total mass that is ejected from the system by the outflow.
The factor $c_{\rm s,g,out}$ is introduced to correct the non-negligible
SNe II contribution in the iron abundance. These are represented as
\begin{eqnarray}
c_{\rm g,s,out} = 1 - 10^{-\left[{\rm O/Fe }\right]_{\rm II}}
\frac{(Z_{\rm O}/Z_{\rm Fe})_{\rm g,s,out}}{(Z_{\rm O}/Z_{\rm Fe})_{\odot}}.
\label{eqn13}
\end{eqnarray}

Using Eq.~({\ref{eqn8}), $r$ is also represented as
\begin{eqnarray}
r = \frac{1}{1+\frac{P_{1,i}}{P_{2,i}} \frac{M_{\rm II}}{M_{\rm
Ia}} \frac{m_{i, \rm Ia}}{m_{i, \rm II}}  \frac{\omega_{\rm
II}}{\omega_{\rm Ia}}\frac{s(u_{\rm now})}{s(u_{\rm now} - \omega t_{\rm Ia}
)}}.
\label{eqn14}
\end{eqnarray}
The $i$ means that $i$th heavy element is used in estimating $r$. We
emphasize that $r$ does not depend on $i$ when results of the
calculation of supernova nucleosynthesis are used. 
This is because $P_{1,i}$
and $P_{2,i}$ are proportional to $m_{i, \rm II}$ and $m_{i, \rm Ia}$, 
respectively (see Eq.~(\ref{eqn71}) and~(\ref{eqn74})).

We can also infer the value $r$ using the present abundance pattern
(Russel \& Dopita 1992) and results of supernova nucleosynthesis as follows
(Tsujimoto et al. 1995). We define the abundance pattern $x_i$ to be
compared with the $x_{i, \rm LMC}$ as
\begin{eqnarray}
x_i(r) = r \frac{m_{i, \rm Ia}}{M_{\rm Ia}} + (1-r)\frac{m_{i, \rm
II}}{M_{\rm II}},
\label{eqn15}
\end{eqnarray}
and the most probable value of $r = r_p$ is determined by minimizing the
following function (Yanagida et al. 1990):
\begin{eqnarray}
g(r) = \sum_{i=1}^{n} [\log x_{i,\rm LMC} - \log x_i(r)]^2/n.
\label{eqn16}
\end{eqnarray}
Tsujimoto et al. (1995) concluded that $r_p = 0.16$ for the LMC from
this fitting (see also Figure 3) using 10 elements (O, Ne, 
Mg, S, Ar, Ca, Cr, Mn, Fe, and Ni).

\subsection{Our Standpoint on Determining the Parameters in the Pagel
Model}\label{extension}
\indent

In this subsection, we explain our standpoint on determining the
parameters in the Pagel Model.

Pagel \& Tautvai$\rm \breve{s}$ien$\rm \dot{e}$ (1998) determined
the parameters in their model so as to explain the age-metallicity
relation, the present mass fraction of gas, and [$\alpha$/Fe] versus
[Fe/H] diagrams in the LMC. $\alpha$ represents $\alpha$-nuclei, such
as O, Mg, Si, Ca, and Ti.

In this study, we add two observational facts, because we should try
to reproduce the observational data as many as possible.
One is the present Fe distribution of stars in the LMC
(Olszewski 1993). It will reflect the SFR history.
The other is the contribution of SNe Ia to the enrichment
of heavy-elements in the gas, $r$.

At first, we can determine the parameters for the gas dynamics, such as 
$g(0)$, $s(0)$, $\omega (t)$, $t_{\rm Ia}$, and $\eta$ as well as
$P_{\rm 1, Fe}$ and $P_{\rm 2,Fe}$ so as to reproduce the observations of the
age-metallicity relation, the present mass fraction of gas, and the
SFR history.

Next, [$\alpha$/Fe] versus [Fe/H] diagrams are used to extract the
contribution of SNe II because $\alpha$-nuclei, especially oxygen,
are synthesized mainly in SNe II (Hashimoto 1995). This means that
$P_{1,\alpha}$ and
$P_{2,\alpha}$ are determined by using the observations of [$\alpha$/Fe]
versus [Fe/H] diagrams. In this study, the [O/Fe] versus
[Fe/H] diagram is mainly used to extract SNe II's
contribution. 
It is also noted that oxygen is synthesized at the outer layer
in a star. So the amount of oxygen in the supernova ejecta is not
influenced so much by the location of the mass cut (the boundary between the
ejecta and the central compact object). On the other hand, most of the 
other $\alpha$-nuclei mentioned above are synthesized near the location
of the mass cut, which means their amounts in the ejecta
are sensitive to the location of the mass cut. That is, the uncertainty 
of the calculations of supernova nucleosynthesis will be smallest when
the amount of oxygen is considered.
Of course, we also discuss the dependence of the results on the
$\alpha$-nuclei to be used in this analysis. This means to check the
validity of the calculation of the supernova nucleosynthesis itself.

At this point, we have to explain the problems with $P_{1,i}$ and $P_{2,i}$.
As is clear from Eq.~(\ref{eqn71}) and~(\ref{eqn74}), $P_{1,i}$
and $P_{2,i}$ are proportional to $m_{i, \rm II}$ and $m_{i, \rm Ia}$, 
respectively. For example, $P_{\rm 1,Fe}$, $P_{\rm 1,O}$,
$P_{\rm 2,Fe}$, and $P_{\rm 2,O}$ meet the relation
\begin{eqnarray}
\frac{P_{\rm 1,O}}{P_{\rm 1,Fe}} = \frac{m_{\rm O, \rm II}}{m_{\rm Fe, \rm II}}
\;\;\;\ \rm and \;\;\; \it \frac{P_{\rm 2,O}}{P_{\rm 2,Fe}} =
\frac{m_{\rm O, \rm Ia}}{m_{\rm Fe, \rm Ia}}. 
\label{eqn10}
\end{eqnarray}
We use the results of calculations for SNe Ia
(Nomoto et al. 1984). So the ratio $P_{\rm 2, O}/ \it P_{\rm 2, Fe}$ is
fixed from the beginning.
In other words, we have only one parameter, $P_{\rm 1,O}$, in
order to fit the [$\alpha$/Fe] versus [Fe/H] diagrams.
If we can not reproduce the observations of the [$\alpha$/Fe] versus
[Fe/H] diagrams completely, it reflects the problem with the
calculation of the nucleosynthesis in a supernova explosion.

However, Pagel and Tautvai$\rm \breve{s}$ien$\rm \dot{e}$ (1998) 
regarded all of $P_{1,i}$ and $P_{2,i}$ to be free parameters.
At the same time, they used Eq.~(\ref{eqn8}) and Eq.~(\ref{eqn92})
in order to estimate the occurrence frequency of SNe Ia relative to
SNe II. It is quite apparent that these treatments are incompatible
with each other. Moreover, they artificially shifted data points upwards
by 0.2 dex for oxygen and downwards by 0.1 dex for silicon so as to 
coincide with the theoretical curves. This means that observational
data points and theoretical curves are shifted freely and we can not
extract useful informations from the chemical evolution model for the LMC.
We emphasize again that we have to determine what is assumed and what
is obtained in the analysis of the chemical evolution model. In this
study, we believe the observational data points and calculation of the
supernova nucleosynthesis. That is, these data points are not
shifted artificially and results of the calculations are not changed
artificially. Parameters in the chemical evolution model are
determined using these informations. This is our standpoint on determining the
parameters in the chemical evolution model.

Finally, the IMF slope is determined from Eq.~(\ref{eqn72}) and
$A$ is determined from Eq.~(\ref{eqn8}) and Eq.~(\ref{eqn81}) using
the results of calculations for SNe II (Hashimoto 1995). 
We can also estimate the occurrence frequency of SNe Ia relative to
SNe II using Eq.~(\ref{eqn8}) and Eq.~(\ref{eqn92}).

\section{ Results} \label{results}
\subsection{Determination of the Parameters}\label{determine}
\indent

Values of the input parameters are tabulated in Table 1.
Those used in the Pagel model and used in this study (New model) are
presented in the table. $P_{1,i}$ and $P_{2,i}$ are
written in units of solar abundance of the corresponding element.
$T_G$ and $t_{\rm Ia}$ are the age of the LMC and the averaged
lifetime of SNe Ia. $\omega$ and time are written in units of
Gyr$^{-1}$ and Gyr, respectively. $g(t=0)/m(t=T_G)$ is the ratio
of the initial mass of gas to the final ($t=T_G$) total baryon mass.
$s(t=0)$ is the initial mass in the form of stars.


The comparison between the theory and the observations is presented in
Figure 1 and Figure 2. 
In Figure 1, theoretical curves of the Pagel model with
observations of the LMC are presented. The age-gas fraction relation,
the age-metallicity relation, the [O/Fe] versus [Fe/H] diagram, and
the present Fe distribution of long-lived stars are shown, respectively.
In Figure 2, theoretical curves with new input parameters are
shown. It is noted that Pagel and Tautvai$\rm \breve{s}$ien$\rm
\dot{e}$ (1998) shifted data points for [O/Fe] upwards by 0.2 dex in
their paper. In this study, such a treatment is not done.
In the [O/Fe] versus [Fe/H] diagram, theoretical curves for $P_{\rm
1,O}/Z_{\odot}^{\rm O}$ = 0.0644 (solid curve) and $P_{\rm
1,O}/Z_{\odot}^{\rm O}$ = 0.154 (short-dashed curve) are shown for
comparison. We can find clearly that solid curve reproduces better the 
[O/Fe] versus [Fe/H] diagram. So we take $P_{\rm 1,O}/Z_{\odot}^{O}$ =
0.0644 as a new parameter.


Values of the output parameters are shown in Table 2.
$\omega_{\rm Ia}$, $\omega_{\rm II}$, $N_{\rm Ia}/N_{\rm II}$, $\dot{N_{\rm
Ia}}/\dot{N_{\rm II}}$, $r$, $x$, and $A$ are shown in the table.
$\chi^2$ is the value of the chi-square probability function when data 
sets for the SFR are used. $\overline{ \rm
|[O/Fe]-[O/Fe]_{\rm obs}|^2}$ is the mean square of the difference
between the observed [O/Fe] and calculated one at the same [Fe/H].
Data points for the [O/Fe] are shifted by 0.2 dex when the value of
the $\overline{ \rm |[O/Fe]-[O/Fe]_{\rm obs}|^2}$ for the Pagel model
is calculated.
As for $r$, the minimizing function $g(r)$ is shown in
Figure 3. $r$ = 0.16 (solid vertical line) is the most
probable value from the fitting of Eq.~(\ref{eqn16}). $r$ = 0.12 and
0.20 are the calculated values by the Pagal and New models, respectively.


Finally, time evolution of the total mass (solid curve), mass of gas
(short-dashed curve), and mass of stars (long-dashed curve) concluded
by the Pagel and New models are presented in Figure 4. Time
is defined as 14 - Age [Gyr].


\subsection{Problem with the Supernova Nucleosynthesis}\label{supernova}
\indent

In this subsection, we compare the theoretical curves for the
[$\alpha$/Fe] versus [Fe/H] diagrams with the observational data points. 
$\alpha$ represents $\alpha$-nuclei, such as Mg, Si, Ca, and Ti.
We can check the validity of the calculation of the supernova
nucleosynthesis due to this test.
We emphasize again that we have no free parameters already, because
$P_{1,i = \rm Mg,Si,Ca,Ti}$ are determined uniquely when $P_{\rm 1,Fe}$
and $P_{\rm 1,O}$ are given.

The results are shown in Figure 5. We can find that 
[Mg/Fe], [Si/Fe], and [Ca/Fe] are not well reproduced as long as
$P_{\rm 1,O}/Z_{\odot}^{\rm O}$ = 0.0644 is adopted.
On the other hand, [O/Fe] is not well reproduced when $P_{\rm
1,O}/Z_{\odot}^{\rm O}$ = 0.154 is adopted (see Figure 2). 
This is the problem what Pagel \& Tautvai$\rm \breve{s}$ien$\rm
\dot{e}$ (1998) pointed out.

As for the ratio of [Ti/Fe], both of the theoretical curves can not
reproduce the data points well. That is, more of Ti is required in
order to explain the observations. This shortage problem of Ti will
mean the shortage problem with the supernova nucleosynthesis
(Tsujimoto et al. 1995; Hashimoto 1995).

Finally, the output parameters for $P_{\rm 1,O}/Z_{\odot}^{\rm O}$ = 0.154
are shown in Table 3. The uncertainties of the output parameters will
be inferred when we compare the values in Table 3 with those in Table 2.


\section{Summary and Discussion} \label{discussion}
\indent

As stated in section 1, chemical evolution model for a
galaxy is a tool from which we can infer the history of the galaxy.
We also emphasize again that a chemical evolution 
model has to contain many parameters in it. We have to determine 
these parameters using results of calculations of supernova 
nucleosynthesis and observations in a galaxy. When these results 
and/or observations are shifted artificially, as done in Pagel \& Tautvai$\rm
\breve{s}$ien$\rm \dot{e}$ (1998), almost all informations are treated
as free parameters. We will be able to extract very little
informations from such works. We must always keep in mind what is
assumed and what is derived in the study of the chemical evolution in
a galaxy. In this study, observations and results of calculations of
supernova nucleosynthesis are believed and used to determine the
parameters in the chemical evolution model. In this section, we
discuss what the New model presented in this study suggests as a result.

As for the ratio $g(t=0)/m(t=T_G)$, the New model requires relatively
high ratio (see Table 1 and Figure 4). This means that
we require an initial condition in which star formation is forbidden
and only gas exists. What can suppress the star formation at zero
metallicity in the LMC? In a simplest situation, a cloud with only
thermal support, collapse should occur if the mass exceeds the Jeans
(1928) mass,
\begin{eqnarray}
M_J = \left( \frac{\pi k T_K}{\mu m_H G} \right) \rho^{-0.5} = 18
M_{\odot} T_K^{1.5} n^{-0.5},
\label{eqn17}
\end{eqnarray}
where $T_K$ is the kinetic temperature (K), $\rho$ is the mass density 
($\rm g \; cm^{-3}$), and $n$ is the total particle density ($\rm
cm^{-3}$). So, low temperature and/or high number density are required 
in order to form a star or, at least, to form a molecular gas cloud.
So, it is suggested that the LMC was born in which the temperature is
high and/or the particle density is low enough to prevent gas from
collapsing and from forming a molecular gas cloud.

We discuss the effect of the outflow. Pagel \& Tautvai$\rm
\breve{s}$ien$\rm \dot{e}$ (1998) introduced an outflow in order to
reduce the metallicity of the
system. The [O/Fe] ratio is also reduced by the effect of the outflow
when a star burst phase is assumed. This is because the material of
SNe II is ejected more than that of SNe Ia. However, the Pagel model
can not reproduce the SFR history very well (see Figure 1 and
the value of the $\chi^2$ probability function in Table 2).
On the other hand, the New model can reproduce the present low metallicity 
and the SFR history at the same time although the New model
requires very smaller value for $\eta$ than the Pagel model. Moreover, 
the [O/Fe] versus [Fe/H] diagram can be reproduced well without
shifting the data points since a low [O/Fe] ratio can be attained by
using a high value of $A$, not by using the effect of the outflow. 
We could not reproduce the observations well when 
$\eta$ is set to be 1.0 like the Pagel model. Although we can not
conclude that our solution is an unique one, one has to construct a new 
parameter set which can reproduce the observations as well as ours if
he insists the existence of the outflow in the LMC. 
Additionally, a steeper IMF slope (Tsujimoto et al. 1995) is not needed
in order to attain the low value of the [O/Fe] since a high value of
$A$ can realize it.

The ratio $N_{\rm Ia}/N_{\rm II}$ inferred from
the New model is very high (=0.48) compared with other models
(Tsujimoto et al. 1995; see also Table 2). This is because high
ratio of $P_{2, \rm Fe}/P_{1, \rm Fe}$ (i.e. high ratio of $A$) is
assumed in the New model (see Table 1). However, $r$=0.20 in
the New model is close to that inferred from the observations (r=0.16; 
see Eq.~(\ref{eqn16})) when the weak dependence of $g(r)$ on $r$
around $r \sim 0.16$ is considered (see Figure 3).
That is why we think the high ratio of $N_{\rm Ia}/N_{\rm II}$ is not
ruled out from the present chemical composition in the LMC.
As for the ratio $\dot{N_{\rm Ia}}/
\dot{N_{\rm II}}$, a higher ratio (= 1.3) is concluded by the New
model than the Pagel model. Since such a high ratio, of order 1, is
estimated by X-ray observations (Hughes et al. 1995), the New model is
thought to be consistent with the observations.

We found that the problems with the supernova nucleosynthesis using
the [$\alpha$/Fe] versus [Fe/H] diagrams. We can not solve these
problems now. However, these problems may be solved if the metallicity
effect for the supernova nucleosynthesis are taken into consideration
(Woosley \& Weaver 1995). We will perform such calculations and report 
its result in the near future.
As for the shortage problem of Ti (Tsujimoto et al. 1995;
Hashimoto 1995), asymmetric explosion models for SNe II may solve the
problem (Nagataki 1997, 1999). This is because more of Ti is produced
in the asymmetric explosion model by the strong alpha--rich freezeout
(Nagataki 1997, 1999).
We will discuss whether these models can explain the [$\alpha$/Fe]
versus [Fe/H] diagrams at the same time in the forth coming paper.

We present a new input parameter set of the Pagel model which
reproduces the observations of the LMC very well. The aim of
constructing a chemical evolution
model is to give some constraints on other theories and numerical
calculations. In order to attain such an aim, the chemical evolution
model has to explain the observations in the galaxy as many as
possible. At the same time, must always keep in mind what is
assumed and what is derived in the study of the chemical evolution in
a galaxy.
Since the parameters in a chemical evolution model are determined by
the observations and results of calculations of supernova
nucleosynthesis, we will be able to determine these parameters more
precisely by further precise observations and/or calculations. 
We hope more
observations and further discussions will be presented by many groups
in the world in order to understand the system of the Large Magellanic
Cloud.

\par
\vspace{1pc}\par
We are grateful to Pagel, B.E.J. for useful comments.
This research has been supported in part by a Grant-in-Aid for the
Center-of-Excellence (COE) Research (07CE2002) and for the Scientific
Research Fund (7449, 199908802) of the Ministry of Education, Science,
Sports and Culture in Japan and by Japan Society for the Promotion of
Science Postdoctoral Fellowships for Research Abroad.

\clearpage
\section*{References}
\re
Barbuy B., de Freitas Pacheco J.A., Castro S.\ 1994, A\&A 283, 32
\re
Geisler D., Bica E., Dottori H., Clar$\rm \acute{i}$a J.J., Piatti
A.E., Santos J.F.C., Jr.\ 1997, AJ 114, 1920 
\re
Girardi L., Chiosi C., Bertelli G., Bressan A.\ 1995, A\&A 298, 87
\re
de Freitas Pacheco J.A., Barbuy B., Idiart T.P.\ 1998, A\&A 332, 19
\re
Hashimoto M.\ 1995, Prog. Theor. Phys. 94, 663
\re
Hill V., Andriewsky S., Spite M.\ 1995, A\&A 293, 347
\re
Hughes J.P., Hayashi I., Helfand D., Hwang U., Itoh M., Kirshner R.,
Koyama K., Markert T., Tsunemi H., Woo J.\ 1995, ApJ 444, L81
\re
Jeans J.H.\ 1928 Astronomy and Cosmogony Cambridge: Cambridge Univ.
Press p.340 
\re
J$\rm \ddot{u}$ttner A., Stahl O., Wolf B., Baschek B.\ 1992, eds B.
Baschek, G. Klare, J. Lequex, New Aspects of Magellanic Cloud
Research. Springer-Verlag, Berlin p.337
\re
Luck R.E., Lambert D.L.\ 1985, ApJ 298, 782
\re
McWilliam A., Williams R.E.\ 1991, eds R. Hayes, D. Milne, The
Magellanic Clouds. Proc. IAU Symp. 148, Kluwer, Dordrecht, p.391
\re
Nagataki S.\ 1999, ApJ 511, 341
\re
Nagataki S.\ 1999, ApJS, accepted, astro-ph/9907109
\re
Nissen P.E., Schuster W.J.\ 1997, A\&A 326, 751
\re
Nomoto K., Thielemann F.-K., Yokoi K.\ 1984, ApJ 471, 903
\re
Olszewski E.W., Schommer R.A., Suntzeff N.B., Harris H.C.\ 1991, AJ
101, 515 
\re
Olszewski E.W.\ 1993, in ASP conference series 48, The Globular
Cluster-Galaxy Connection, eds. G.H. Smith and J.P. Brodie (A.S.P:San
Francisco) p.351  
\re
Pagel B.E.J., Tautvai$\rm \breve{s}$ien$\rm \dot{e}$ G.\ 1995, MNRAS
276, 505 
\re
Pagel B.E.J., Tautvai$\rm \breve{s}$ien$\rm \dot{e}$ G.\ 1998, MNRAS
299, 535 
\re
Richtler T., Spite M., Spite F.\ 1989, A\&A 225, 351
\re
Russel S.C., Dopita M.A.\ 1992, ApJ 384, 508
\re
Salpeter E.E.\ 1955, ApJ 121, 161
\re
Spite M., Barbuy B., Spite F.\ 1989, A\&A 222, 35
\re
Tsujimoto T., Nomoto K., Yoshii Y., Hashimoto M., Yanagida S.,
Thielemann F.-K.\ 1995, MNRAS 277, 945
\re
Westerlund B.\ 1997, The Magellanic Clouds, Cambridge University Press.
\re
Woosley S.E., Weaver T.A.\ 1995, ApJS 101, 181
\re
Yanagida S., Nomoto K., Hayakawa S.\ 1990, Proceedings of the 21st
International Cosmic Ray Conference, Adelaide 4, 44
\re
\clearpage

\begin{table*}
\begin{center}
Table~1. \hspace{4pt} Values of the Input Parameters.\\
\end{center}
\vspace{6pt}
\begin{tabular*}{\textwidth}{@{\hspace{\tabcolsep}
\extracolsep{\fill}}p{6pc}|llllllll}
\hline \hline\\[-6pt]
 & Pagel Model & New Model \\
\hline
$P_{1, \rm Fe}/Z_{\odot}^{\rm Fe}$ & 0.28 & 0.07 \\
$P_{1, \rm O}/Z_{\odot}^{\rm O}$   & 0.70 & 0.0644 \\
$P_{2, \rm Fe}/Z_{\odot}^{\rm Fe}$ & 0.42 & 0.25 \\
$T_G$ (Gyr)       & 14   & 14 \\
$t_{\rm Ia}$ (Gyr)& 1.33 & 2.2 \\
$\eta$            & 1.0  & 0.001 \\
$\omega (t)$      & 0.15 $\; (t \le 2)$         & 0.35 $\; (t \le 0.2)$\\
                  & 0.08 $\; (2 \le t \le 11)$  & 0.12 $\; (0.2 \le t
\le 11)$\\
                  & 0.50 $\; (11 \le t \le 14)$ & 0.45 $\; (11 \le t
\le 14)$\\  
$g(0)/m(T_G)$        & 0 & 0.82 \\
$s(t = 0)$               & 0 & 0   \\ 
\hline
\end{tabular*}
\vspace{6pt}\par\noindent
Left column: those used in the Pagel
model. Right: those used in this study. $P_{1,i}$ and $P_{2,i}$ are
written in units of solar abundance of the corresponding element.
$T_G$ and $t_{\rm Ia}$ are the age of the LMC and the averaged
lifetime of SNe Ia. $\omega$ and time are written in units of
Gyr$^{-1}$ and Gyr, respectively. $g(t=0)/m(t=T_G)$ is the ratio
of the initial mass of gas to the final ($t=T_G$) total baryon mass.
$s(t=0)$ is the initial mass in the form of stars.
\end{table*}

\begin{table*}
\begin{center}
Table~2. \hspace{4pt} Values of the Output Parameters.\\
\end{center}
\vspace{6pt}
\begin{tabular*}{\textwidth}{@{\hspace{\tabcolsep}
\extracolsep{\fill}}p{10pc}|lllllll}
\hline \hline\\[-6pt]
 & Pagel Model & New Model \\
\hline
$\omega_{\rm Ia}$                            & 0.21 & 0.25 \\
$\omega_{\rm II}$                            & 0.14 & 0.21 \\
$N_{\rm Ia}/N_{\rm II}$                      & 0.20 & 0.48 \\
$\dot{N_{\rm Ia}}/\dot{N_{\rm II}}$          & 0.42 & 1.3  \\
$r$                                          & 0.12 & 0.20 \\
$x$                                          & 1.01 & 1.06 \\
$A$                                          & 0.075& 0.17 \\
$\chi^2$                                     & 0.41 & 0.73 \\
$\overline{\rm |[O/Fe]-[O/Fe]_{\rm obs}|^2}$ & 0.19 & 0.19 \\
\hline
\end{tabular*}
\vspace{6pt}\par\noindent
Left column: those derived by the Pagel
model. Right: those derived by this study. $\omega_{\rm Ia}$ and
$\omega_{\rm II}$ are mass fractions of heavy-element ejected into the
interstellar gas from SNe Ia and SNe II, respectively. $N_{\rm
Ia}/N_{\rm II}$ and $\dot{N_{\rm Ia}}/\dot{N_{\rm II}}$ are the
occurrence frequency of SNe Ia relative to SNe II and the ratio of the current
supernova rates, respectively. $r$ is the contribution of SNe Ia to the
enrichment of heavy-elements in the gas. $x$ is the IMF slope. $A$ is
the probability for 3-8$M_{\odot}$ stars to explode as SNe Ia.
$\chi^2$ is the value of the chi-square probability function when data 
sets for the star formation rate are used.
$\overline{ \rm |[O/Fe]-[O/Fe]_{\rm obs}|^2}$ is the mean square of the
difference between the observed [O/Fe] and calculated one at the
same [Fe/H].  
\end{table*}

\begin{table*}
\begin{center}
Table~3. \hspace{4pt} Values of the Output Parameters for $P_{\rm
1,O}/Z_{\odot}^{O}$ = 0.154.\\
\end{center}
\vspace{6pt}
\begin{tabular*}{\textwidth}{@{\hspace{\tabcolsep}
\extracolsep{\fill}}p{10pc}|lllllll}
\hline \hline\\[-6pt]
$\omega_{\rm Ia}$                            &  0.25 \\
$\omega_{\rm II}$                            &  0.21 \\
$N_{\rm Ia}/N_{\rm II}$                      &  0.43 \\
$\dot{N_{\rm Ia}}/\dot{N_{\rm II}}$          &  1.16  \\
$r$                                          &  0.13 \\
$x$                                          &  0.22 \\
$A$                                          &  0.51 \\
$\chi^2$                                     &  0.73 \\
$\overline{\rm |[O/Fe]-[O/Fe]_{\rm obs}|^2}$ &  0.40 \\
\hline
\end{tabular*}
\vspace{6pt}\par\noindent
Same as Table 2 but for $P_{\rm 1,O}/Z_{\odot}^{\rm O}$ = 0.154 model.
\end{table*}

\clearpage
\centerline{Figure Captions}
\bigskip
\begin{fv}{1}
{7cm}
{
Theoretical curves of the Pagel model with observations of the LMC.
The upper panel on the left: the age-gas fraction relation. Data point 
is from Westerlund (1997).
The lower panel on the left: the age-metallicity relation.
Data sources are as follows: $open \; triangles$, Olszewski $et \;
al$. (1991); $filled \; triangles$, Girardi $et \; al$. (1995);
$open \; hexagons$, Geisler $et \; al$. (1997);
$open \; squares$, de Freitas Pacheco et al. (1998). 
The upper panel 
on the right: the [O/Fe] versus [Fe/H] diagram. Data sources are as
follows: $open \; triangles$, Luck \& Lambert (1992);
$open \; squares$, Th$\rm \acute{e}$venin (1997);
$open \; hexagons$, Hill, Andriewski \& Spite (1995);
$filled \; triangles$, Spite, Barbuy \& Spite (1993);
$filled \; hexagons$, J$\rm \ddot{u}$tter $et \; al$. (1992);
$five-pointed \; stars$, McWilliam \& Williams (1991);
$three-pointed \; stars$, Barbuy, de Freitas Pacheco \& Castro (1994);
$cross$, Richtler, Spite \& Spite (1989);
$asterisks$, Nissen \& Schuster (1997).  
The lower panel on the right: the present Fe distribution of
long-lived stars. Data source is from Olszewski (1993).
}
\end{fv}
\begin{fv}{2}
{7cm}
{Same as Fig.1 but with new input parameters.
In the [O/Fe] versus [Fe/H] diagram, theoretical curves for $P_{\rm
1,O}/Z_{\odot}^{\rm O}$ = 0.0644 (solid curve) and $P_{\rm
1,O}/Z_{\odot}^{\rm O}$ = 0.154 (short-dashed curve) are shown for
comparison. 
}
\end{fv}
\begin{fv}{3}
{7cm}
{The minimizing function $g(r)$ as a function of r. 10 elements (O, Ne, 
Mg, S, Ar, Ca, Cr, Mn, Fe, and Ni) are used for the fitting. $r$ =
0.16 (solid vertical line) is the most probable value. $r$ = 0.12 and
0.20 are the calculated values by the Pagal and New models, respectively.
}
\end{fv}
\begin{fv}{4}
{7cm}
{Time evolution of the total mass (solid curve), mass of gas
(short-dashed curve), and mass of stars (long-dashed curve). Time is
defined as 14 - Age [Gyr]. 
}
\end{fv}
\begin{fv}{5}
{7cm}
{[$\alpha$/Fe] versus [Fe/H] diagrams with the observational data
points. Solid curves represent the theoretical curves for $P_{\rm
1,O}/Z_{\odot}^{\rm O}$ = 0.0644. Short-dashed curves represent the ones
for $P_{\rm 1,O}/Z_{\odot}^{\rm O}$ = 0.154.
}
\end{fv}

\thispagestyle{empty}
\begin{figure}
\begin{center}
   \leavevmode\psfig{figure=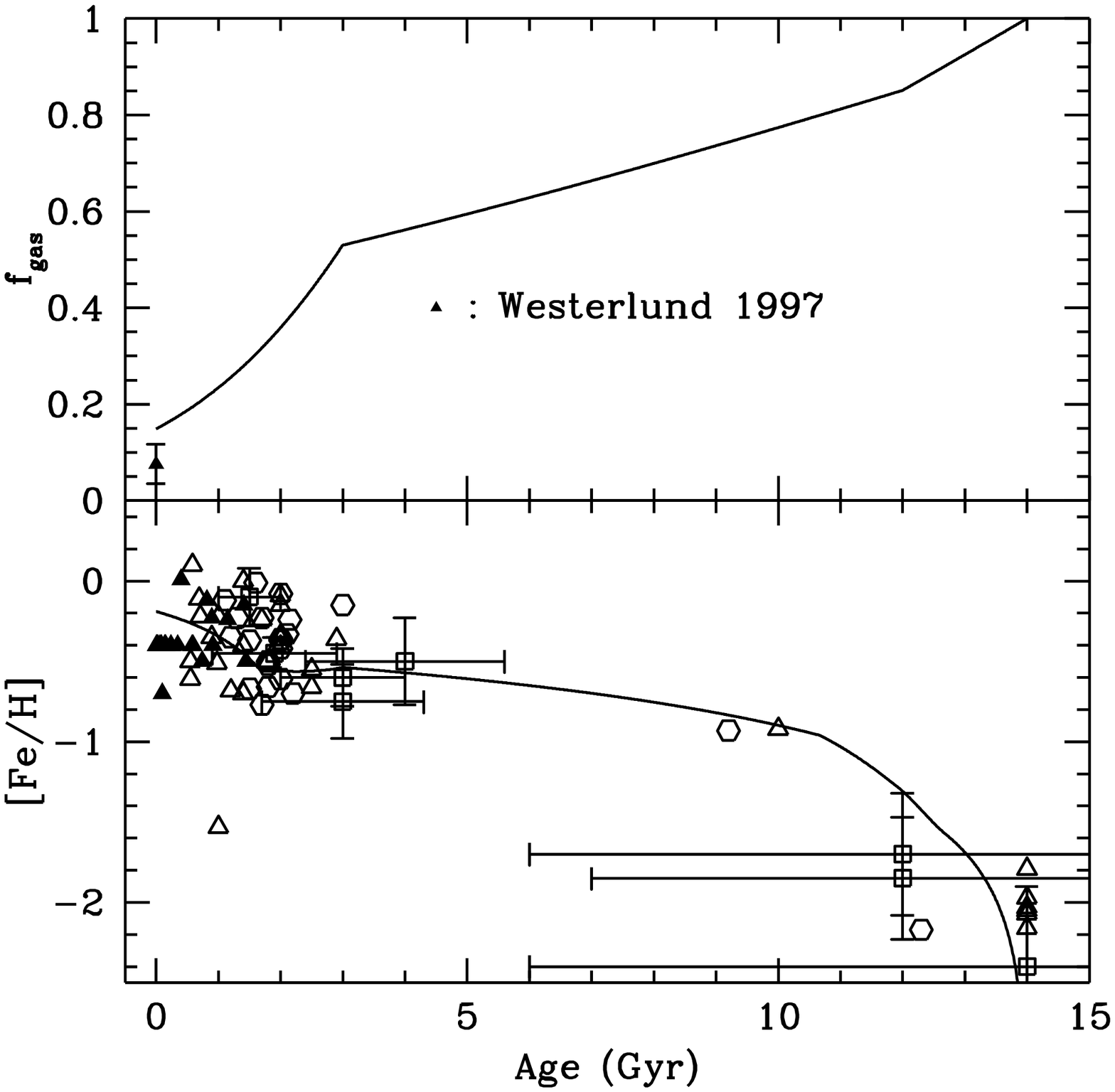,height=7cm,angle=0}
   \leavevmode\psfig{figure=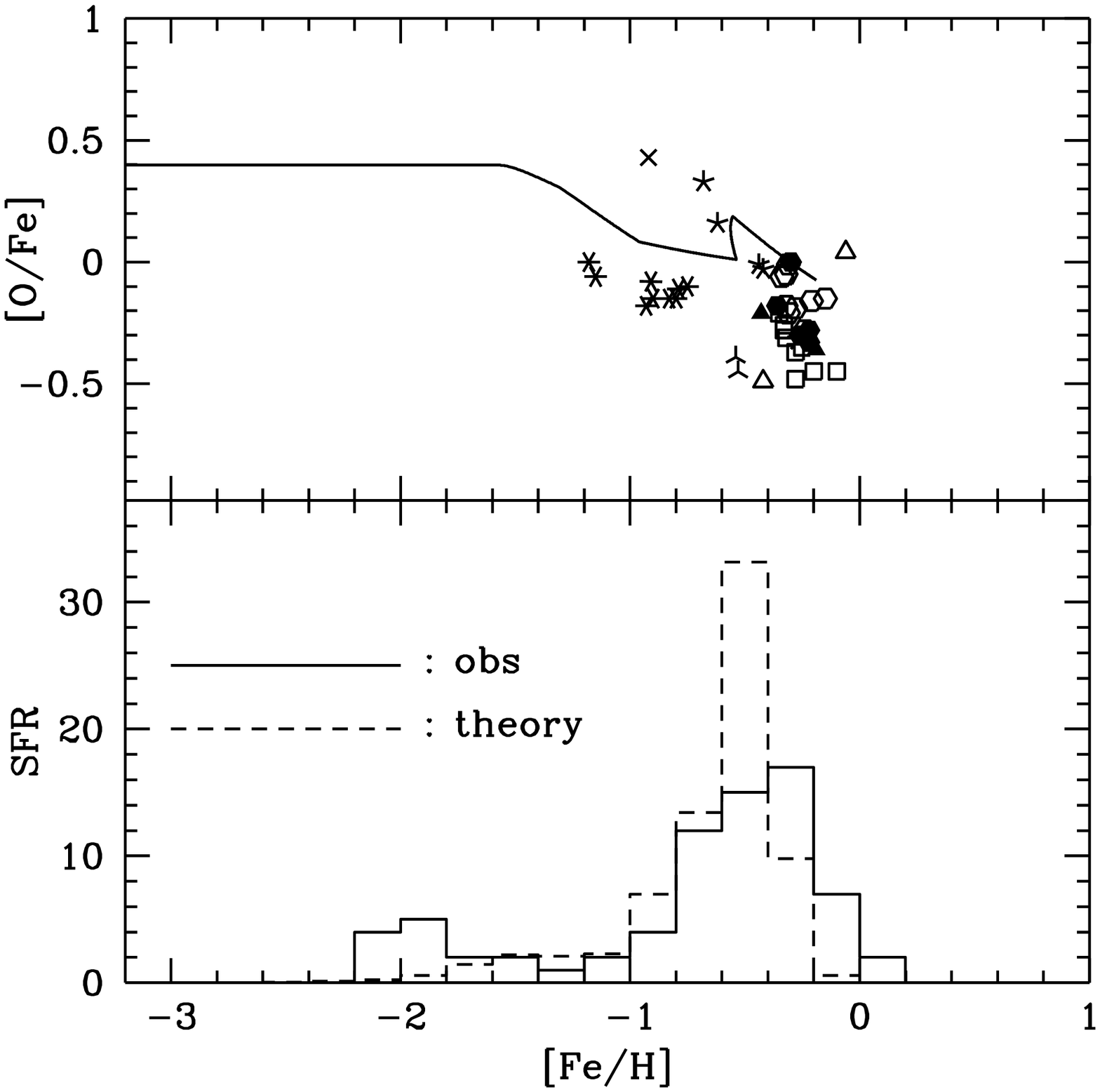,height=7cm,angle=0}
\end{center}
\caption{}
\label{fig1}
\end{figure}

\thispagestyle{empty}
\begin{figure}
\begin{center}
   \leavevmode\psfig{figure=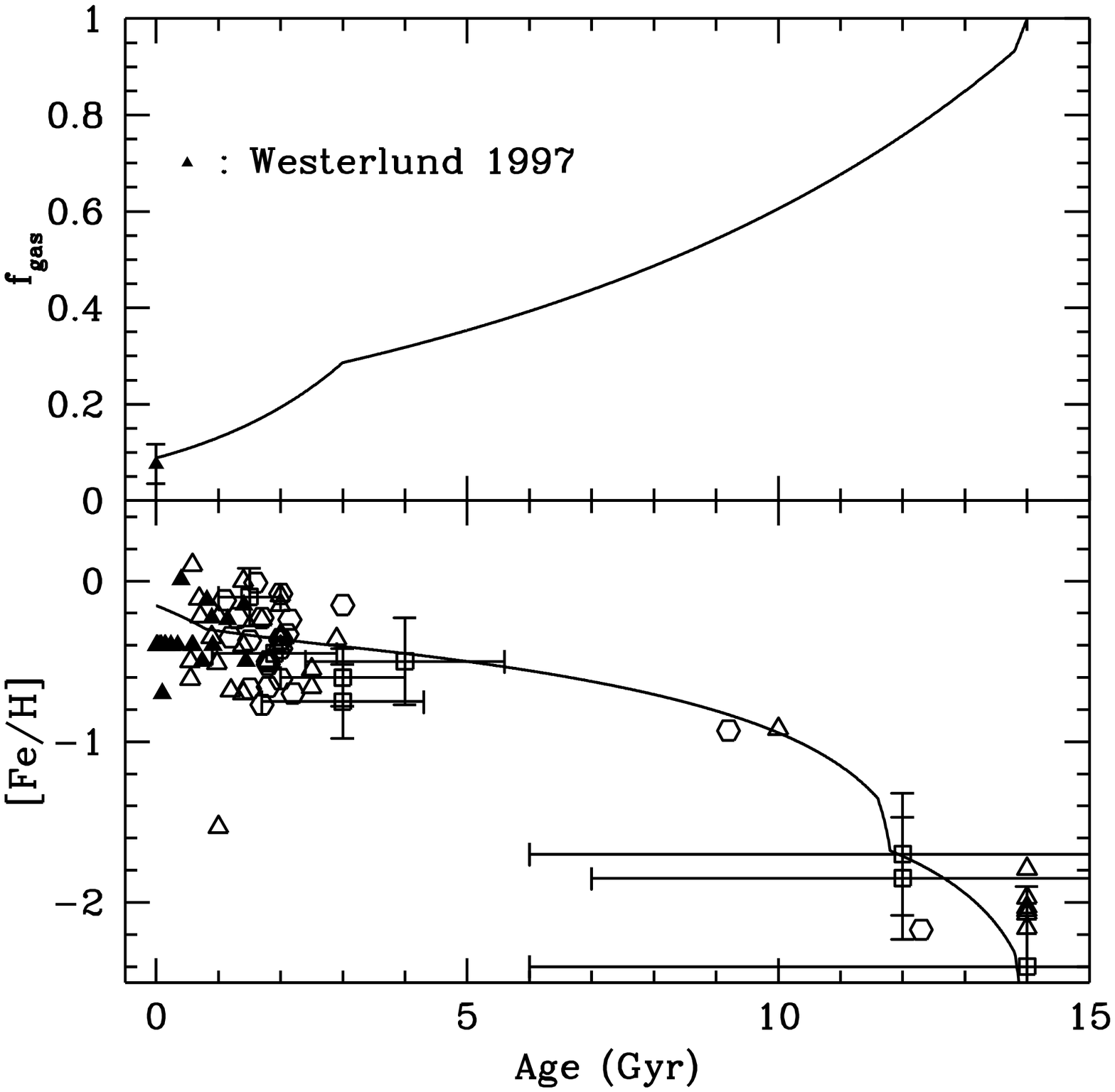,height=7cm,angle=0}
   \leavevmode\psfig{figure=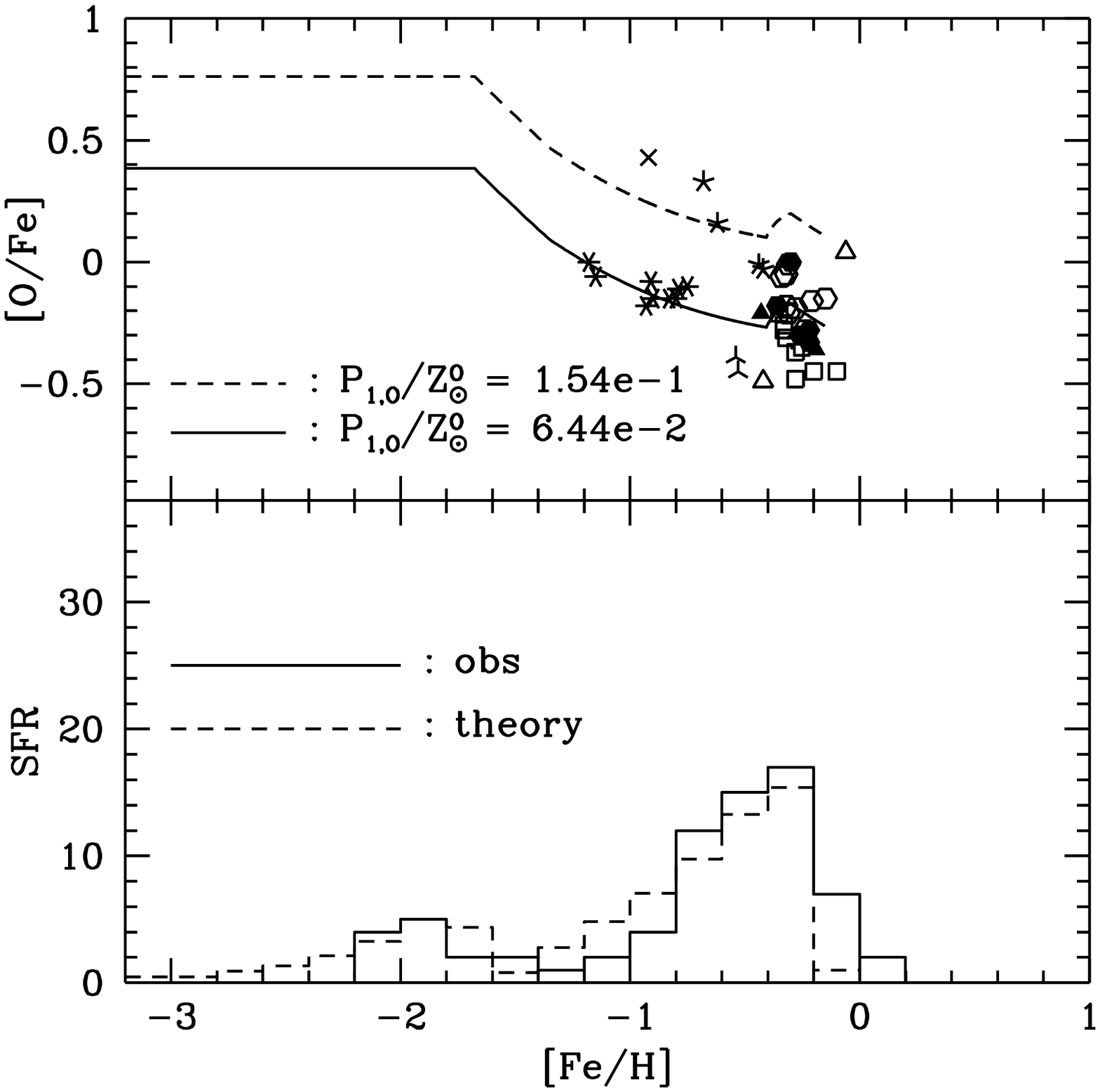,height=7cm,angle=0}
\end{center}
\caption{}
\label{fig2}
\end{figure}

\thispagestyle{empty}
\begin{figure}
\begin{center}
   \leavevmode\psfig{figure=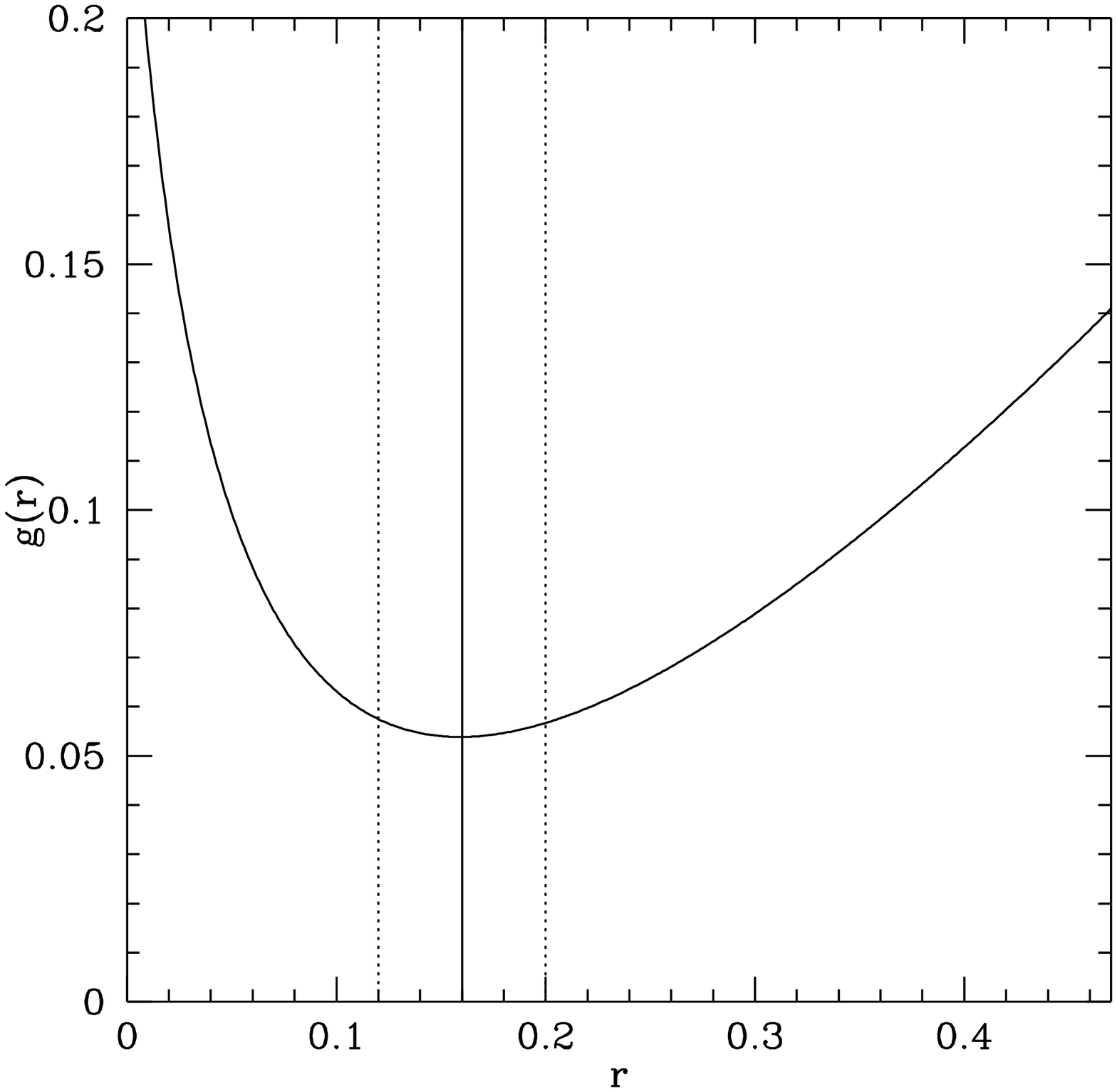,height=7cm,angle=0}
\end{center}
\caption{}
\label{fig3}
\end{figure}

\thispagestyle{empty}
\begin{figure}
\begin{center}
   \leavevmode\psfig{figure=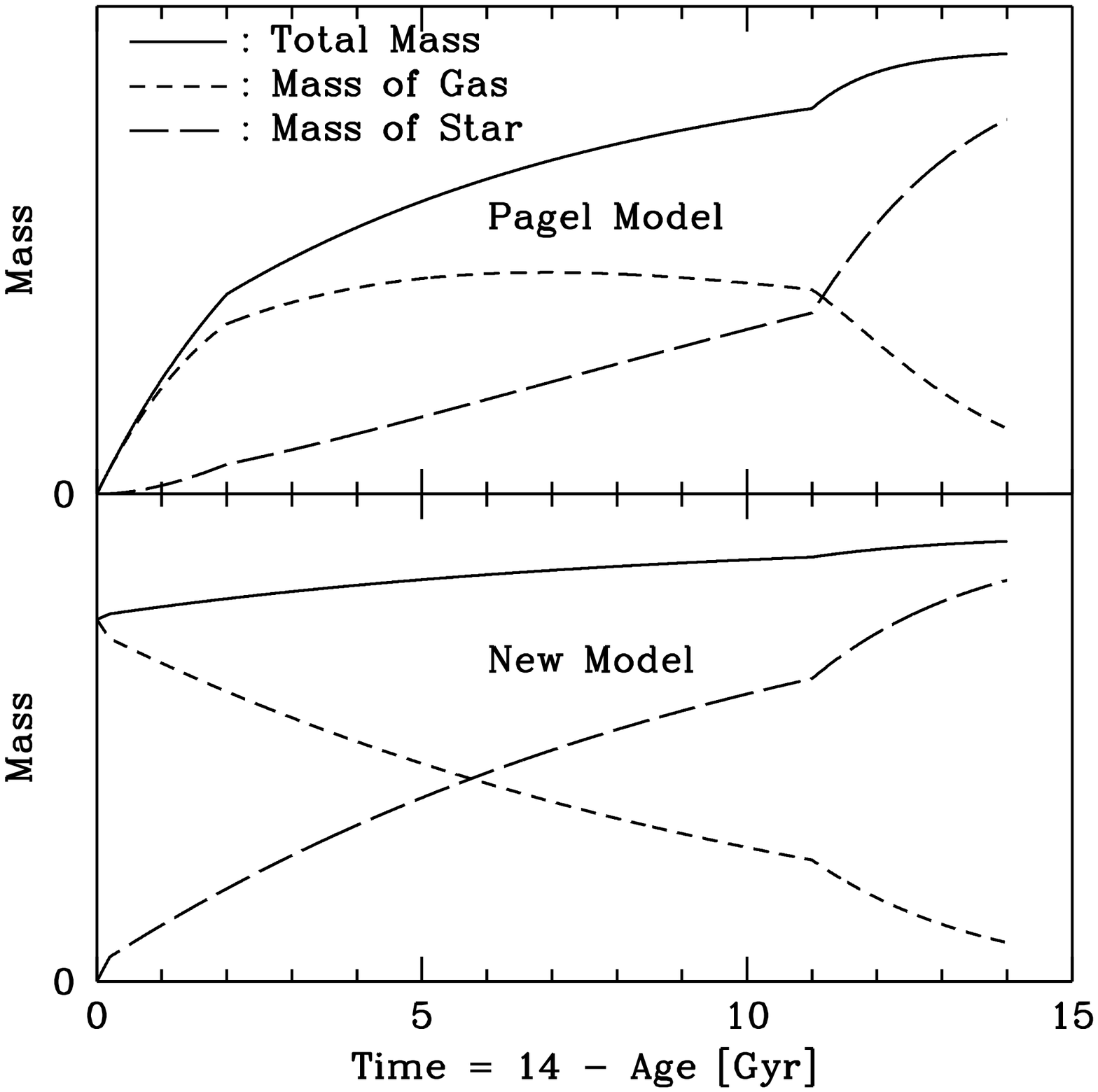,height=7cm,angle=0}
\end{center}
\caption{}
\label{fig4}
\end{figure}

\thispagestyle{empty}
\begin{figure}
\begin{center}
   \leavevmode\psfig{figure=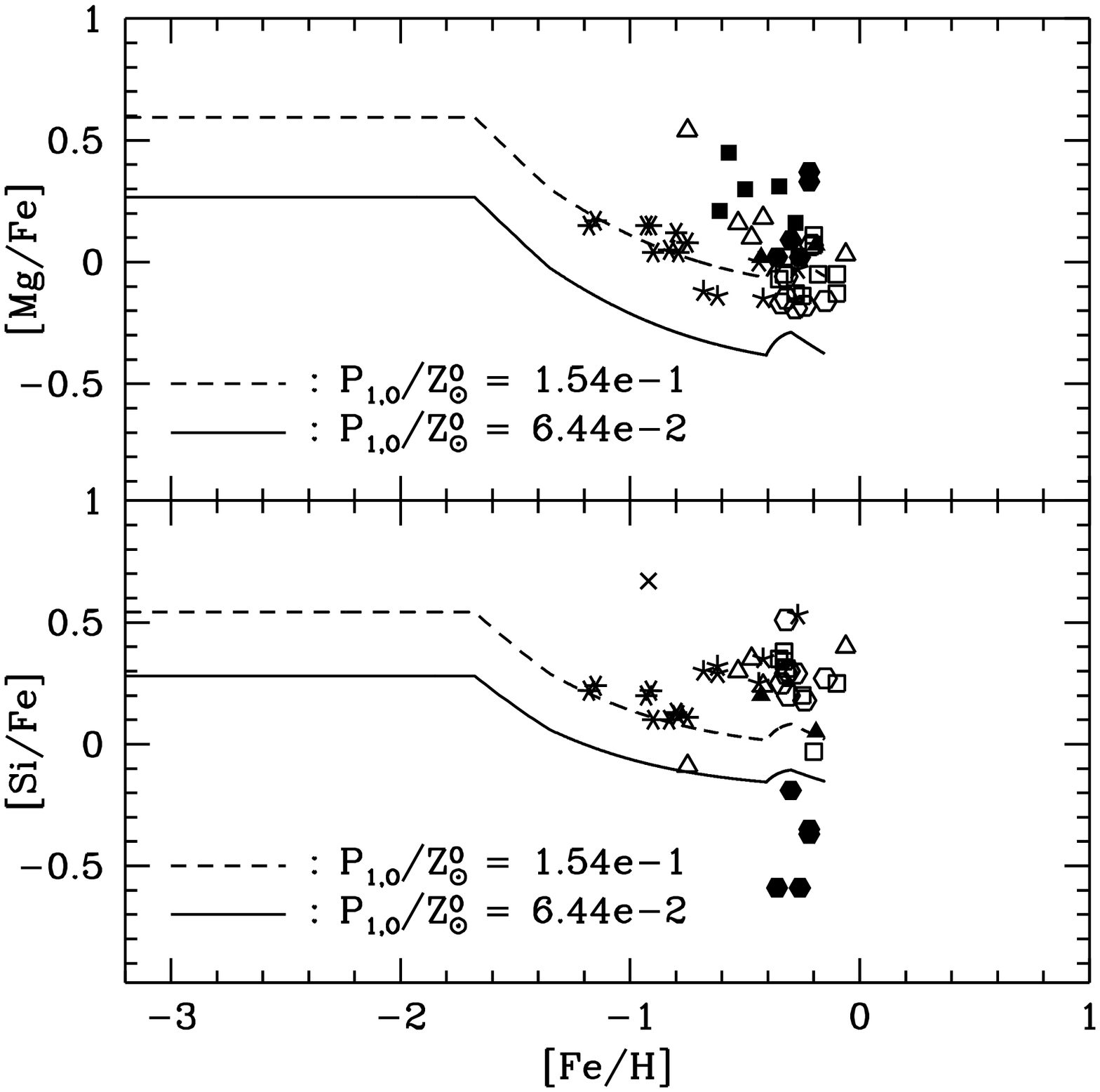,height=7cm,angle=0}
   \leavevmode\psfig{figure=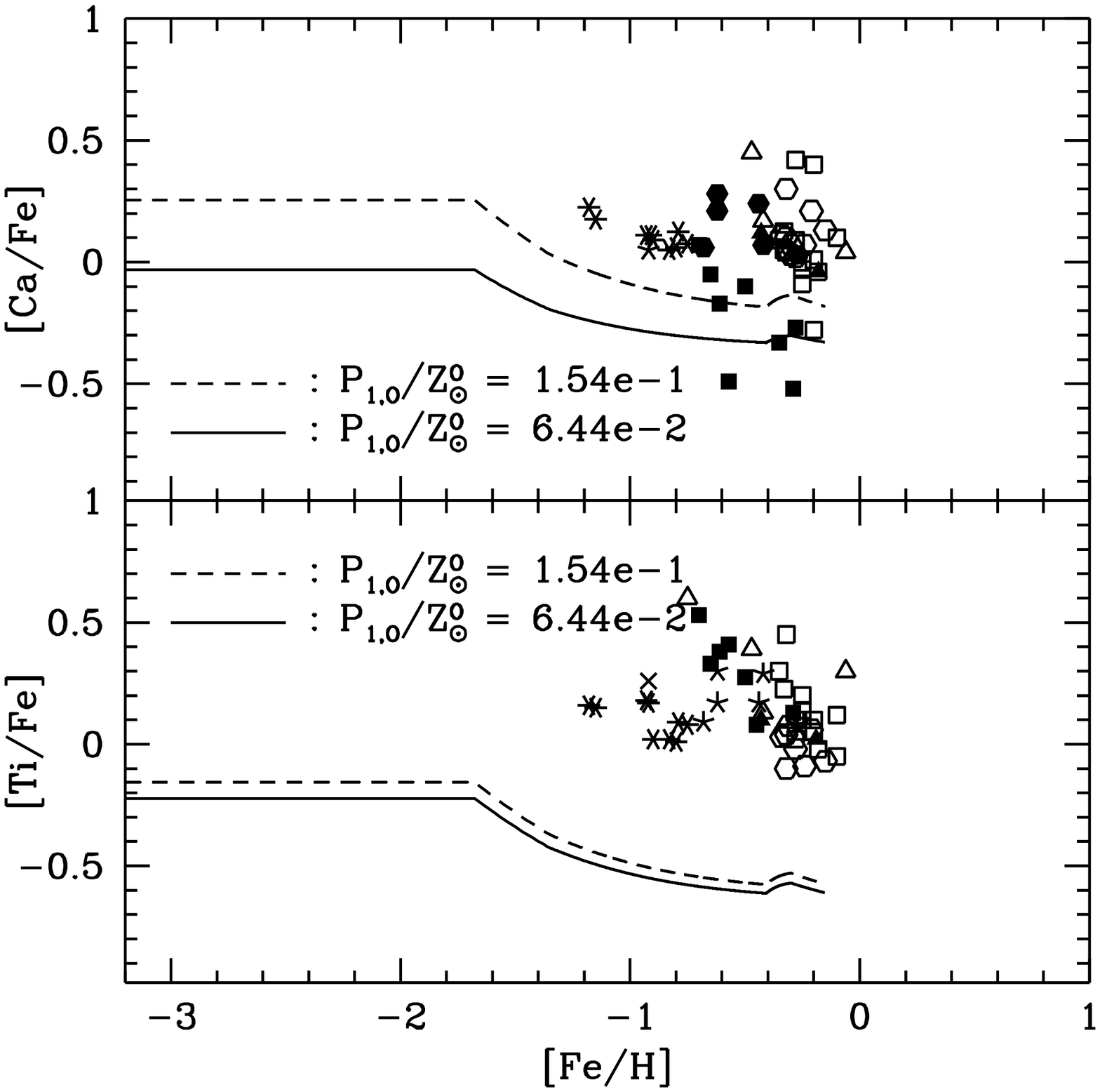,height=7cm,angle=0}
\end{center}
\caption{}
\label{fig5}
\end{figure}

\end{document}